\documentstyle[aps,epsf]{revtex}  

\newcommand{\ra}    {\rightarrow}
\newcommand{\babar} {{\it B}{\footnotesize \it A}{\it B}{\footnotesize \it AR}}

\begin{document}       

\title{ \hfill \hbox{\normalsize DE-AC02-76SF00515 \qquad DE-FG02-91ER-40622 
        \qquad SLAC-PUB-10756} \break \break
\hbox{\normalsize DPF 2003: Annual Meeting of the Division of Particles
and Fields of the American Physical Society} \break
\hbox{\normalsize 5-8 April 2003, Philadelphia, Pennsylvania} \break
 \break
Study of $B$ Decays into the Final State $D^{(*)}\pi\pi\pi$ at $\babar$}

\author{H. W. Zhao}
\address{University of Mississippi--Oxford}
\maketitle

\begin{abstract}        

Based on a data sample of an integrated luminosity of 57.4 fb$^{-1}$
at the $\Upsilon (4S)$ resonance
taken with the \babar \ detector using the SLAC PEP-II
asymmetric-energy $B$-Factory, hadronic decays of $B$ meson with the final 
states $D^{(*)}$ and three pions are studied. The study is performed by fully 
reconstructing the exclusive decays of $B \ra D^{(*)} a_{1}(1260)$ and 
the non-resonant modes $B \ra D^{(*)}\rho^{0}\pi$ and 
$B \ra D^{(*)}\pi\pi\pi$. The current status of the study is presented and
a dominant $B \ra D^{(*)} a_{1}(1260)$ decay is shown. 

\end{abstract}

\section{Introduction}              

Hadronic $B$ decays provide important information on both the weak and 
hadronic interactions of heavy flavored mesons. The quarks produced in such 
nonleptonic weak decays can arrange themselves into hadrons in many ways. 
The final state is linked to the initial state by QCD processes. 
The theoretical description of hadronic decays of heavy mesons invoke the 
factorization approximation and heavy quark effective theory (HQET) [1].\\

The dominant hadronic decay modes of the $B$ meson involve tree-level 
diagrams where the $b \ra c$ transition leads to a charmed meson and an 
external $W$ boson, which often emerges as a charged meson $\pi$, $\rho$ or
$a_{1}(1260)$. In the models based on the Bauer-Stech-Wirbel 
approach [1,2], two parameters $a_{1}$ and $a_{2}$, which describe the QCD 
hard-gluon corrections for external and internal spectator processes 
respectively, provide important clues into the role played by the strong 
interaction in these two-body tree-level decays. Parameters $a_{1}$ and 
$a_{2}$ are expected to be process dependent, but previous experimental data 
can be described with universal values ($a_{1} \approx 1$ and 
$a_{2} \approx 0.2$). Recent results of $B^{0} 
\ra \bar{D}^{(*)0}\pi^{0}$ from CLEO and Belle (both indicate $a_{2}$ to be
$\approx 0.4$) begin 
to show the expected process dependence of $a_{2}$ [3,4]. A precise 
measurement of $B$ decaying to $D^{(*)}a_{1}(1260)$ provides an interesting 
approach to estimate the constants $a_{1}$ and $a_{2}$ and to test the 
factorization 
hypothesis. Since the QCD interaction between the product quarks
continues after the weak transition takes place and after hadron formation, 
an understanding of final state interactions (FSI) is very important. 
The study of the non-resonant modes 
$B \ra D^{(*)}\rho^{0}\pi$ and $B \ra D^{(*)}\pi\pi\pi$ may provide 
important information on this mechanism. \\

Hadronic $B$ decays also play a significant role in CP violation study.
A precise measurement of $B^{0} \ra D^{(*)-}a_{1}^{+}$ is important in 
measuring $\sin(2\beta + \gamma )$, where $\beta$ and $\gamma$ are the 
angles of the Unitary Triangle related to the $CP$ asymmetries in $B$ decay
[5]. Presently $B^{0} \ra D^{*-}a_{1}^{+}$ is also suggested as 
a test of chirality [6].\\

Studies of $B \ra D^{(*)}a_{1}$ and of their non-resonant decays were 
previously performed by CLEO and ARGUS [7,8,9,10]. Because of limited 
statistics, the uncertainties of the branching fraction measurements 
are relatively large [11], and some of the detailed substructure of the
resonance is not available. \babar \ has collected a large data sample of 
$B\bar{B}$ pairs at the PEP-II $B$-Factory [12], making possible an improved 
study of these modes. \\

In this paper we present the status of our study. The \babar \ detector and 
data are described in Section II, and the event selection with data analysis 
is presented in Section III. A summary is given in Section IV. Charge 
conjugation is always implied in this report.

\section{THE \babar \ DETECTOR AND DATA SET}

The data used in this analysis consists of 62.2 million $B\bar{B}$ pairs,
corresponding to an integrated luminosity of 57.4 fb$^{-1}$, collected with
the \babar \ detector at the $\Upsilon(4S)$ resonance.
The \babar \ detector [13] was constructed to observe  CP violation and
was motivated by the measurements of $B^0$ mixing made almost two decades
ago [14].
Charged particles
are reconstructed with a five layer, double-sided silicon vertex tracker (SVT)
[15]
and a 40 layer drift chamber (DCH) with a helium-based gas mixture [16], placed
in a 1.5 T solenoidal field produced by a superconducting magnet. The charged
particle momentum resolution is approximately $(\delta p_{T})^{2} =
(0.0013p_{T})^{2} + (0.0045)^{2}$, where $p_{T}$ is given in GeV/$c$. The
SVT, with a typical single-hit resolution of 10 $\mu$m, provides measurement
of impact parameters of charged particle tracks in both the plane transverse
to the beam direction and along the beam. Charged particle types are
identified from the ionization energy loss (d$E$/d$x$) measured in the DCH
and SVT, and the Cherenkov device (DIRC) [17].
Photons are identified by a CsI(Tl)
electromagnetic calorimeter (EMC) with an energy resolution $\sigma(E)/E =
0.023 \cdot (E/$GeV)$^{-1/4} \oplus 0.019$ [18].

\section{EVENT SELECTION AND DATA ANALYSIS}

$B$ mesons are reconstructed in modes of $B$ decaying to $D^{(*)}$ and three 
pions, which may form the resonance $a_{1}(1260)$ and non-resonant 
$\rho^{0}\pi$ and $\pi\pi\pi$. They contain 
$B^{0} \ra D^{*-} a_{1}^{+}$, $B^{+} \ra {\bar D^{*0}} a_{1}^{+}$, 
$B^{0} \ra D^{-} a_{1}^{+}$, $B^{+} \ra {\bar D^{0}} a_{1}^{+}$, and their
corresponding non-resonant modes $D^{(*)}\rho^{0}\pi$ and $D^{(*)}\pi\pi\pi$.
\\

To reconstruct the charmed mesons $D$ and $D^{*}$ and the light mesons in the 
final states, the charged tracks are required to have a distance of closest 
approach within $\pm 10$ cm in $z$ and 1.5 cm in radius of the average beam 
spot position, and at least 12 hits recorded in the DCH. Kaons and pions are 
selected, depending on their track's momentum, based on the d$E$/d$x$ 
information from the SVT and DCH, as well as the Cherenkov angle and the 
number of photons measured with the DIRC. For each detector component 
$d$ = (SVT, DCH, DIRC), a likelihood $L_{d}^{K}$ ($L_{d}^{\pi}$) is 
calculated given the kaon (pion) mass hypothesis. Photons are selected as 
showers in the calorimeter that have a lateral shower shape consistent with 
the expected pattern of energy deposits for an electromagnetic shower and 
with an energy $E_{\gamma}>30$ MeV.\\

The $D^{0}$ meson is selected from the decays $K^{-}\pi^{+}$, 
$K^{-}\pi^{+}\pi^{0}$ and $K^{-}\pi^{+}\pi^{-}\pi^{+}$, while the
$D^{+}$ meson is selected from $K^{-}\pi^{+}\pi^{+}$ and 
$K_{S}^{0}\pi^{+}$ modes. The track momentum is required to be greater 
than 200 MeV/c. $\pi^{0}$s are reconstructed from two photons with the sum 
of their energies $E_{\gamma \gamma}$ greater than 200 MeV/c and an invariant 
mass 0.120 GeV/c$^{2}$ $<m_{\gamma\gamma}<$ 0.150 GeV/c$^{2}$ (about 
$\pm 2.5$ times the resolution of the $\pi^{0}$ invariant mass). A mass 
constraint fit is applied to $\pi^{0}$ candidates, which improves the energy 
resolution from 2.4\% to 1.8\%. $K_{S}^{0}$ mesons are reconstructed from 
two pion tracks with opposite charge. The invariant mass of the candidate 
$\pi^{+}\pi^{-}$ is required to be $|m_{\pi\pi}-497.7$ MeV/c$^{2}$ 
$| < 15.0$ MeV/c$^{2}$ and the momentum of the $K_{S}^{0}$ greater than 200 
MeV. In $K_{S}^{0}$ selection 
the angle between the flight direction, which is the vector which points from 
the primary vertex to the $\pi^{+}\pi^{-}$ candidate's vertex, and the 
direction of the $\pi^{+}\pi^{-}$ candidate's three momentum is used as a cut.
A vertex fit is applied, and a probability of $\chi^{2}$ greater than 
0.1\% is required. 
The invariant mass of the candidate of $D^{0} \ra K^{-}\pi^{+}$, 
$K^{-}\pi^{+}\pi^{-}\pi^{+}$, and $D^{+} \ra K^{-}\pi^{+}\pi^{+}$, 
$K^{0}_{S}\pi^{+}$ is required to be within $3 \sigma$ from the true value 
of the mass. As the combinatorial background of the 
$D^{0} \ra K^{-}\pi^{+}\pi^{0}$ is larger due to the presence of the 
$\pi^{0}$, a $2 \sigma$ mass cut is applied on this mode. The tracks of the 
$D$ meson are required to originate from the same point; therefore a vertex 
fit is applied with the requirement of a probability of $\chi^{2}$ greater 
than 0.1\%. Then a combined vertex and mass constraint fit is applied and a 
convergent fit is required. The mass-constraint fit changes the momenta 
of the $D$ meson daughters forcing the invariant mass to be the nominal one, 
thus improving the $D$ energy and momentum resolutions.\\

Reconstructed $D^{0}$s are combined with soft pions $\pi^{-}$ ($\pi^{0}$)
to form $D^{*-}$ ($D^{*0}$) candidates. The soft pion momentum is required to
be less than 450 MeV/c. The requirements of $E_{\gamma \gamma}>200$ 
MeV for $\pi^{0}$ and the minimum momentum requirement of 200 MeV/c for the 
pion are removed for soft pions. $D^{*}$ candidates are selected by the 
requirement that the mass difference between $D^{*}$ and $D^{0}$, 
$\Delta m = m_{D^{*}}- m_{D^{0}}$, lies within $\pm 3 \sigma$ ($\sigma$ is 
the resolution of $\Delta m$) of the nominal mass difference. Then 
$D^{*+} \ra  D^{0}\pi^{+}$ candidates are refitted with the beam-spot 
constraint to improve the angular resolution for the soft pion,
and a convergent combined vertex and mass constraint fit is also applied.
For $D^{*0} \ra D^{0}\pi^{0}$, no vertex fit is applied, but a kinematic and 
mass constraint fit are applied and required to be convergent. \\

$\rho^{0}$ mesons are reconstructed from $\pi^{+}\pi^{-}$ pairs 
with the requirement of pion momenta greater than 200 MeV/c and the invariant 
mass satisfying
$|M_{\pi^{+}\pi^{-}} - 0.770$ GeV/c$^{2}$ $| < 0.15 $ GeV/c$^{2}$. 
$a^{\pm}_{1}$ mesons are reconstructed by combining the selected 
$\rho^{0}$ and a pion with the pion momentum greater than 200 MeV/c and the 
invariant mass $m_{\rho^{0}\pi^{\pm}}$ to be between 1.0 GeV/c$^{2}$ and 1.6 
GeV/c$^{2}$. A vertex fit is applied to the candidates of $\rho^{0}$ and 
$a^{\pm}_{1}$, and a $\chi^{2}$ probability greater than 0.1\% is 
required. \\

The $B^{0}$ and $B^{+}$ mesons are reconstructed by combining selected 
$D$ or $D^{*}$ with $a_{1}$ candidates, $\rho^{0}\pi$ or $\pi\pi\pi$. 
The variables $\Delta E$ and $m_{ES}$ are used to define the $B$ signal. 
$\Delta E = E^{*}_{B} - E^{*}_{beam}$ is the energy difference between 
the energy of the $B$ candidate and the beam energy in the $\Upsilon(4S)$ 
system, i.e., $E^{*}_{B}$ is the center of mass energy of the $B$ candidate 
and $E^{*}_{beam}$ is the center of mass beam energy. The $B$ signal is 
expected to peak at $|\Delta E| = 0$. The beam energy substituted mass of 
$B$ candidate, 
$m_{ES}$, is defined as $m_{ES} = \sqrt{E_{beam}^{2} - (\sum_{i} 
\vec p^{\,*}_{i})^{2}}$ where $\vec p^{\,*}_{i}$ is the center of mass 
momentum of 
the $i$-th daughter of the $B$ candidate. The resolution in this variable 
is limited by the beam energy spread, which is about 2.7 MeV for \babar.  
Since the final states involve many tracks and more than one candidate can be 
found in the event, the combinatorial background is high. To suppress such 
background, we chose for each mode the $B$ 
candidates whose daughter $D$ or $D^{*}$ masses are most consistent with 
their nominal given masses. Then among them the candidate with smallest 
$|\Delta E|$ is selected. This selection will remove most of the 
combinatorial background and keep the best candidates.\\ 

One major source of background comes from the $q\bar{q}$ continuum background.
The continuum background is due to the large non-resonant fraction of the 
hadronic cross-section, approximately 75\% at the  
$\Upsilon(4S)$, from direct $e^{+}e^{-} \ra q\bar{q}~~~(q=c,s,u,d)$. To select 
$B{\bar B}$ events from $\Upsilon(4S)$ data and reduce the continuum 
background, the ratio of second to zeroth Fox-Wolfram moment, $R_{2}$, is 
used [19]. In the rest frame of $\Upsilon(4S)$, $R_{2}$ approaches zero for 
spherical events and one for jet-like events (Fig. \,1). $R_{2} < 0.4$ is 
required in event selection since the momenta of $B$'s are very small 
and $B{\bar B}$ events are spherical, whereas $q\bar{q}$ continuum events are
jet-like in the frame of $\Upsilon(4S)$.

\begin{figure}[ht]
\centering
\tabcolsep=6mm
\begin{tabular}{cc}
\begin{minipage}[t]{2.8in}
\centerline{\epsfxsize 2 truein \epsfbox{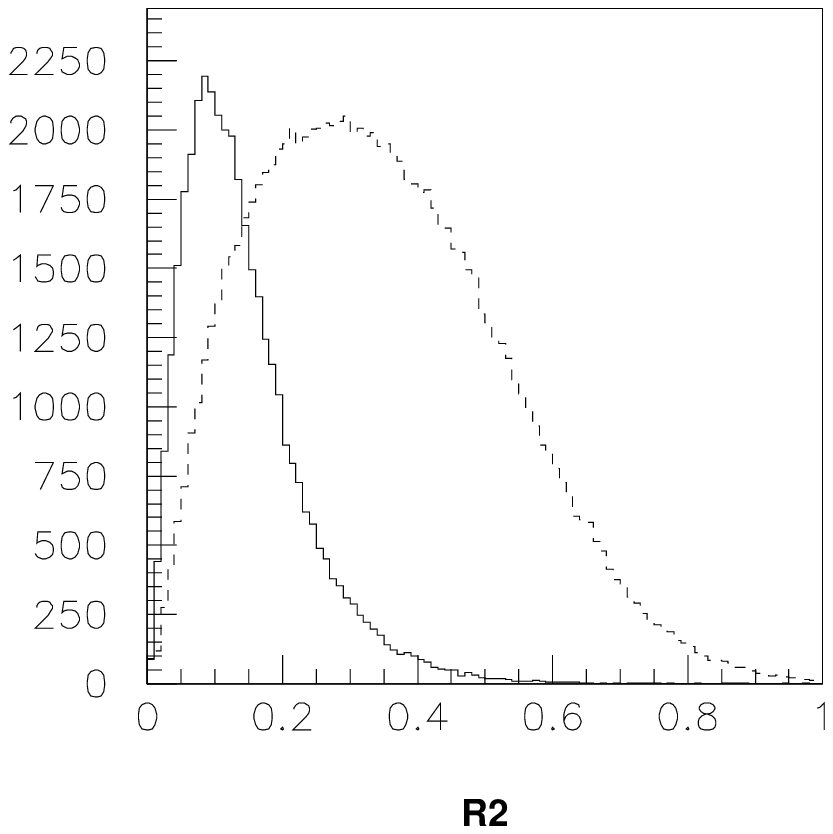}}
\vskip 0.3 cm
\caption[]{
\label{LegsFigure1}
\small Distribution of $R_{2}$. The solid-line distribution is for
$B\bar{B}$ Monte Carlo events and the dashed-line distribution is for
continuum $q\bar{q}$ Monte Carlo events.}
\end{minipage}

&

\begin{minipage}[t]{2.8in}
\centerline{\epsfxsize 2 truein \epsfbox{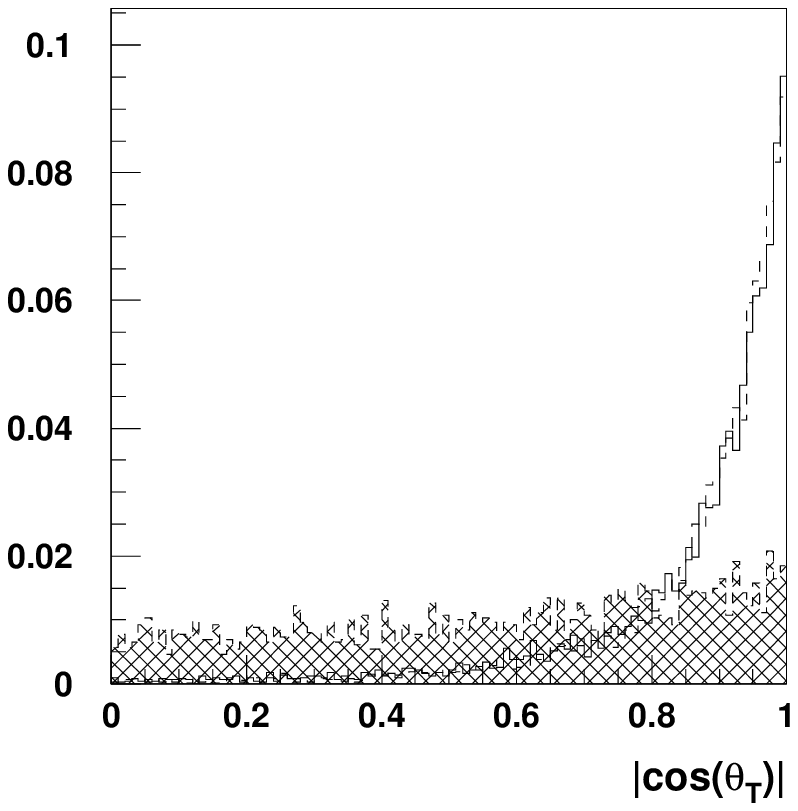}}
\vskip 0.3 cm
\caption[]{
\label{LegsFigure2}
\small Normalized distribution of $|\cos(\theta_{T})|$.
The hatched area is the distribution for $B\bar{B}$ Monte Carlo events.
The solid-line and dashed-line histograms are for continuum $c\bar{c}$ and
$q\bar{q}$ ($q=u,d,s$) Monte Carlo events respectively. (The vertical axis 
has arbitrary units).}
\end{minipage}\\
\end{tabular}
\end{figure}

The thrust angle $\theta_{T}$, which is the angle between the thrust 
axis of the $B$ candidate and the thrust axis of the remaining tracks in 
the event, is also used to suppress the continuum background. Monte Carlo 
simulation shows that the distribution of the thrust angle for 
continuum background events and of $B\bar{B}$ events are quite different 
(Fig. \,2). The distribution of $|\cos(\theta_{T})|$ is flat for 
$B\bar{B}$ events and peaks at 1.0 for continuum $q\bar{q}$ events. 
We require $|\cos(\theta_{T})| < 0.70$ for the $B^{-} \ra D^{0} a_{1}^{-}$, 
$D^{0} \rho^{0}\pi^{-}$ and $D^{0} \pi^{+}\pi^{-}\pi^{-}$ modes with 
$|cos(\theta_{T})| <0.85$ for all other modes. \\

\subsection{ Resonant mode $B^{(*)} \ra D^{(*)} a_{1}(1260)$ }

In this decay $B$ mesons are selected in the modes 
$B^{0} \ra D^{*-} a_{1}^{+}$, $B^{+} \ra {\bar D^{*0}} a_{1}^{+}$, 
$B^{0} \ra D^{-} a_{1}^{+}$ and $B^{+} \ra {\bar D^{0}} a_{1}^{+}$. 
The $a^{+}_{1}$ meson is a very broad ($\Gamma \approx 400$ MeV) 
isovector ($I=1$) state with $l=1$ orbital excitation and $J^{P} = 1^{+}$. 
Fig. \,3 shows the Monte Carlo simulated spectra of $a^{+}_{1}$ mass and 
momenta in the decay of $B \ra D^{(*)} a_{1}(1260)$. 
$a^{+}_{1}$ are reconstructed from a combination of selected $\rho^{0}$ 
and a charged pion, with the pion momentum greater than 200 MeV/c and the 
invariant mass $m_{\rho^{0}\pi^{\pm}}$ between 1.0 GeV/c$^{2}$ and 1.6 
GeV/c$^{2}$ consistent with the $a^{+}_{1}$ mass.  An additional cut on 
the center of mass momentum of the $a_{1}$ candidate is applied with 
$p^{*}_{a_{1}} > 0.5 $ GeV/c.
A vertex fit is performed  and a $\chi^{2}$ probability greater than 
0.1\% is required. All selected $D^{0}$ and $D^{+}$ candidates from $B$ 
decays or $D^{*}$ decays are required to have a momentum $p^{*}_{D}$ 
in the $\Upsilon(4S)$ frame greater than 1.3 GeV/c. The $\Delta E$ 
distributions for $B \ra D^{(*)} a_{1}(1260)$ with $m_{ES} > 5.27$ GeV/c$^{2}$
are shown in Fig. \,4, while the $m_{ES}$ distributions 
with $|\Delta E|< 2.5 \sigma_{\Delta E}$ are shown in Fig. \,5 and Fig. \,6. 

\begin{figure}[ht]
\centerline{\epsfxsize 5.0 truein \epsfbox{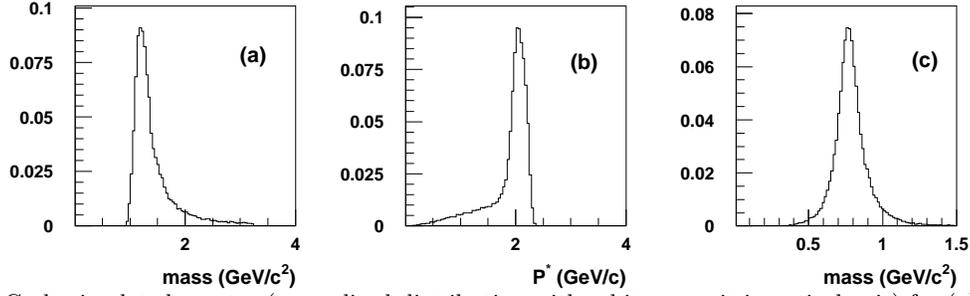}}   
\vskip -.0 cm
\caption[]{
\label{LegsFigure3}
\small  Monte Carlo simulated spectra (normalized distribution with arbitrary 
unit invertical axis) for (a) $a^{+}_{1}$ mass and (b) momenta 
in the decay of $B \ra D^{(*)} a_{1}(1260)$, and (c) the mass of $\rho^{0}$
from $a_{1}$.}
\end{figure}

The $m_{ES}$ background can be separated into continuum and $B\bar{B}$ 
components. The $B\bar{B}$ background component is the result of 
mis-reconstructing other $B\bar{B}$ decays. The relative contributions and 
the overall amount of background varies decay-mode by decay-mode depending 
primarily on the multiplicity of the $B$ decay. We use generic Monte Carlo 
$B\bar{B}$ data with the signal modes $B \ra D^{(*)} a_{1}$ and non-resonant 
modes $B \ra D^{(*)}\rho^{0}\pi$ and $B \ra D^{(*)}\pi\pi\pi$ removed,
and continuum Monte Carlo $q\bar{q}$ data to model the background and find that
the background shape of data can be well characterized by Monte Carlo 
(Fig. \,7). The Monte Carlo backgrounds are fitted to an Argus function [20]
and the obtained shape parameters of such functions are used in the fitting of 
$m_{ES}$ of data as shown in Fig. \,5 and Fig. \,6.

\begin{figure}[ht]
\centerline{\epsfxsize 3.5 truein \epsfbox{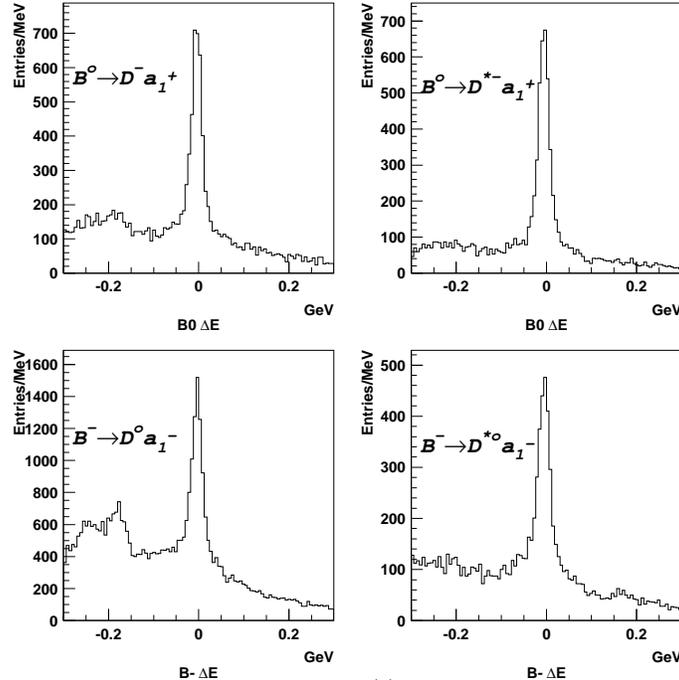}}   
\vskip -.1 cm
\caption[]{
\label{LegsFigure4}
\small 
$\Delta E$ distributions for $B^{0} \ra D^{(*)-}a_{1}^{+}$
with $m_{ES}>5.27$ MeV/c$^{2}$.}
\end{figure}

In the signal region of $B \ra D^{(*)}a_{1}$, some events of the non-resonant 
decays of $B \ra D^{(*)}\rho^{0}\pi$ and $B \ra D^{(*)}\pi\pi\pi$ can 
pass the event selection criteria of $B \ra D^{(*)}a_{1}$ and form a peaking 
background. This is confirmed by $B\bar{B}$ Monte Carlo data. Fig. \,8 shows 
that peaking background of all the non-resonant modes underlays the Monte 
Carlo signal of $B \ra D^{(*)}a_{1}$. The estimation of contributions from
non-resonant modes to $B \ra D^{(*)}a_{1}$ modes will be given in the next 
subsection. It is also possible that the decay of $B^{0} 
\ra D^{*-} \pi^{+}\pi^{+}$ may pollute the signal of $B^{-} \ra D^{0} 
a_{1}^{-}$ by misreconstructing $\rho^{0}$ from one direct $\pi^{+}$ and the 
slow $\pi^{-}$ from $D^{*-}$, where the mass of $\rho^{0}\pi^{+}$ lies in 
the $a_{1}$ region. Monte Carlo study shows that compared with the decays 
$B \ra D^{*} \rho^{0} \pi^{-}$ and  $B \ra D^{*} \pi^{-} \pi^{+} \pi^{-}$,
its contribution is quite small and can be neglected. Another possible 
source of background is from the decay 
$B^{-} \ra D_{1}^{0}(2420) \pi^{-}~(\rho^{-})$ and 
$B^{-} \ra D_{2}^{0}(2460) \pi^{-}~(\rho^{-})$, where $D_{1}^{0}$ or 
$D_{2}^{0}$ decays to $D^{*+} \pi^{-}$ [21]. Monte Carlo study also shows that
their contributions are quite small and can be neglected.

\begin{figure}[ht]
\centerline{
\epsfxsize 2.0 truein \epsfbox{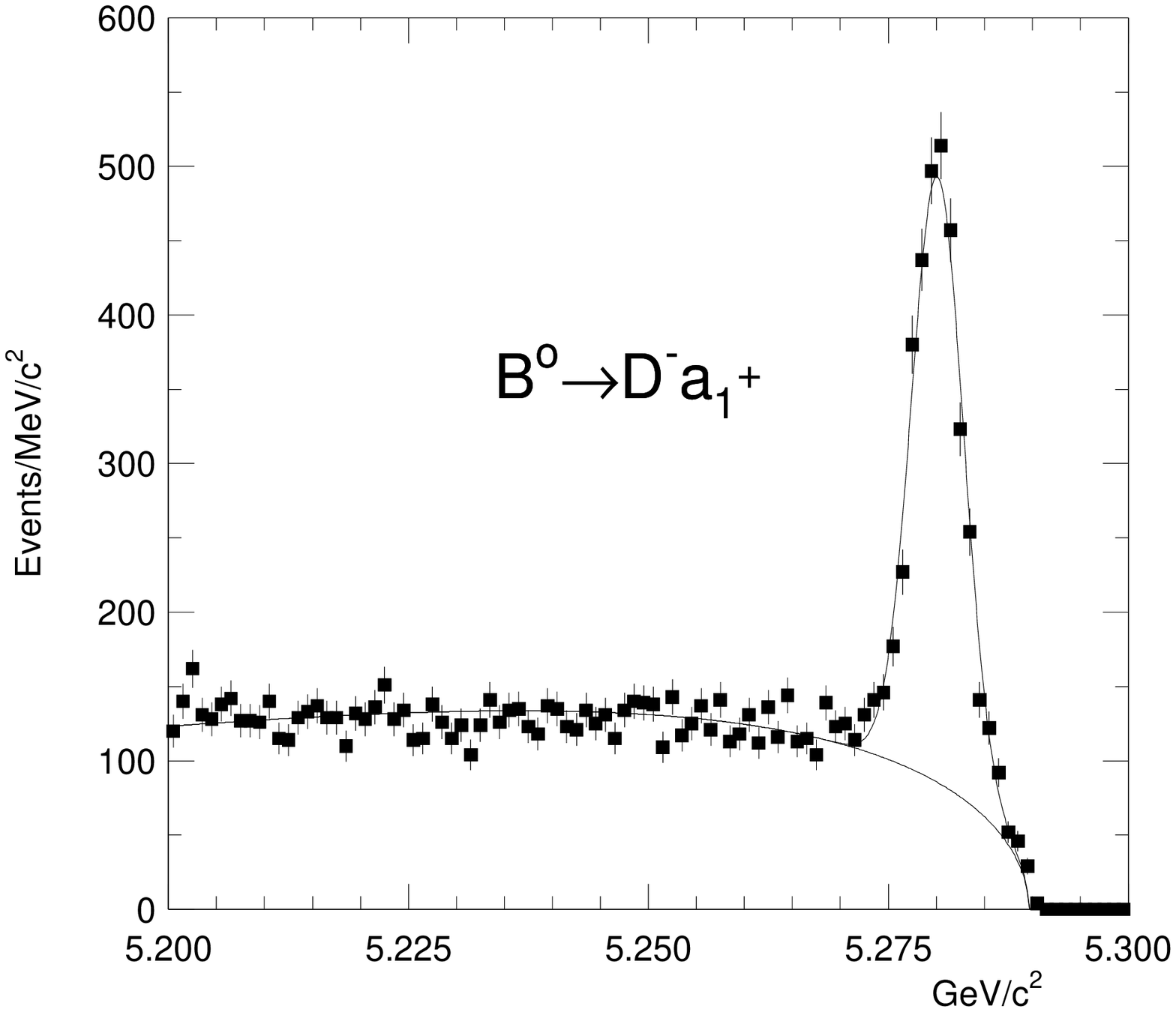}
\epsfxsize 2.0 truein \epsfbox{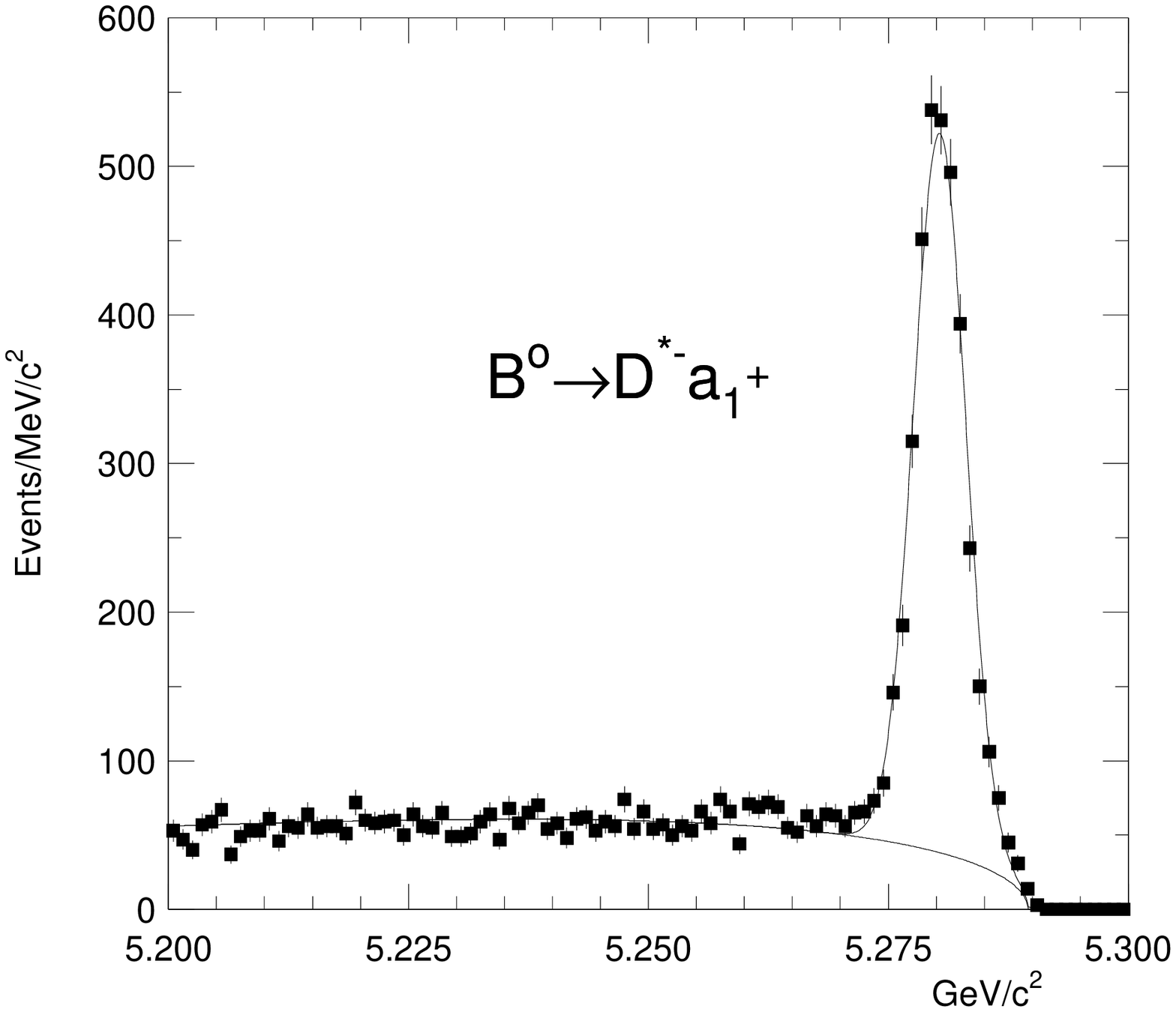}}      
\caption[]{
\label{LegsFigure5}
\small $m_{ES}$ distributions for $B^{0} \ra D^{(*)-}a_{1}^{+}$
with $|\Delta E|< 2.5 \sigma_{\Delta E}$.}
\end{figure}

\begin{figure}[ht]
\centerline{
\epsfxsize 2.0 truein \epsfbox{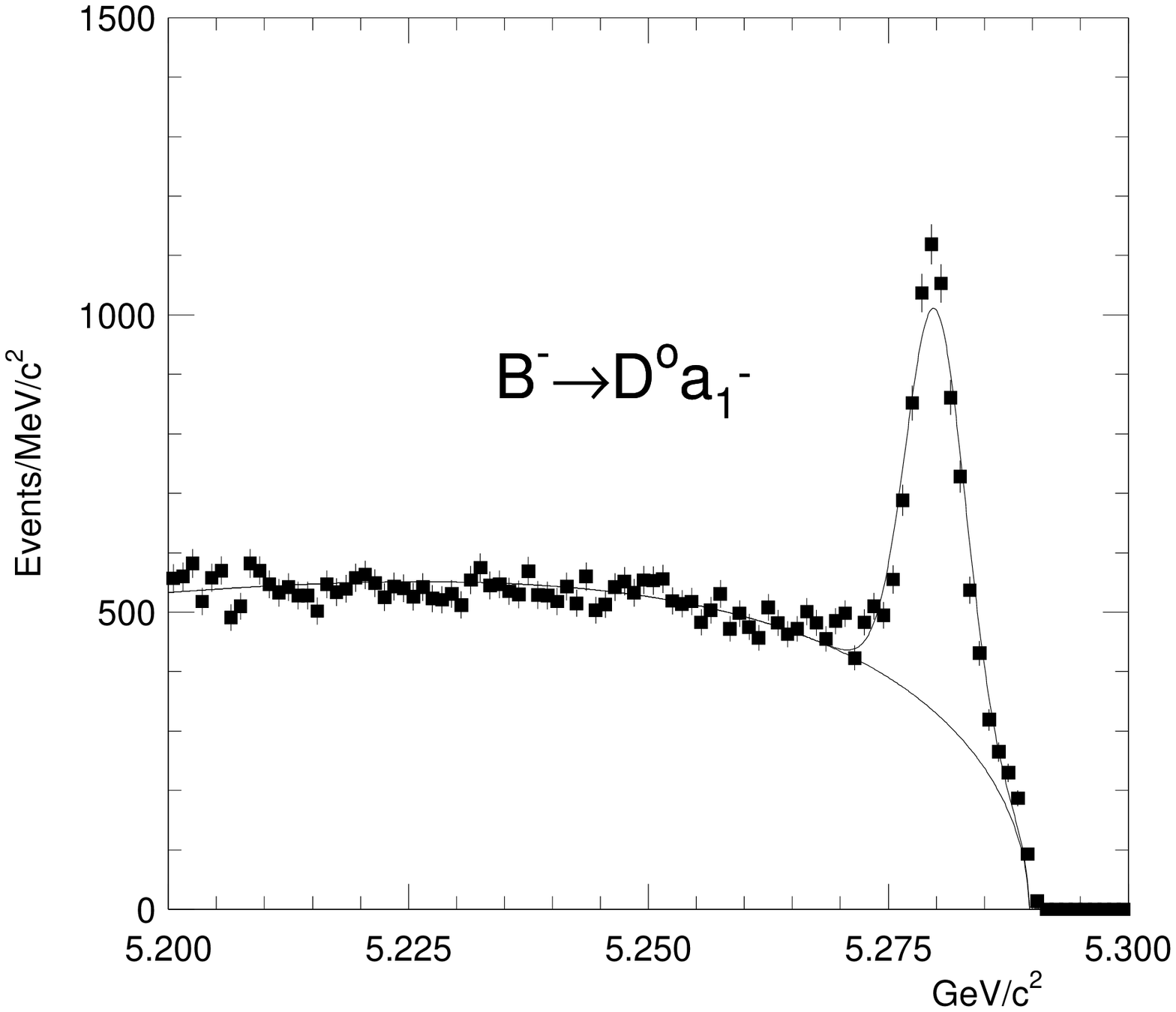}
\epsfxsize 2.0 truein \epsfbox{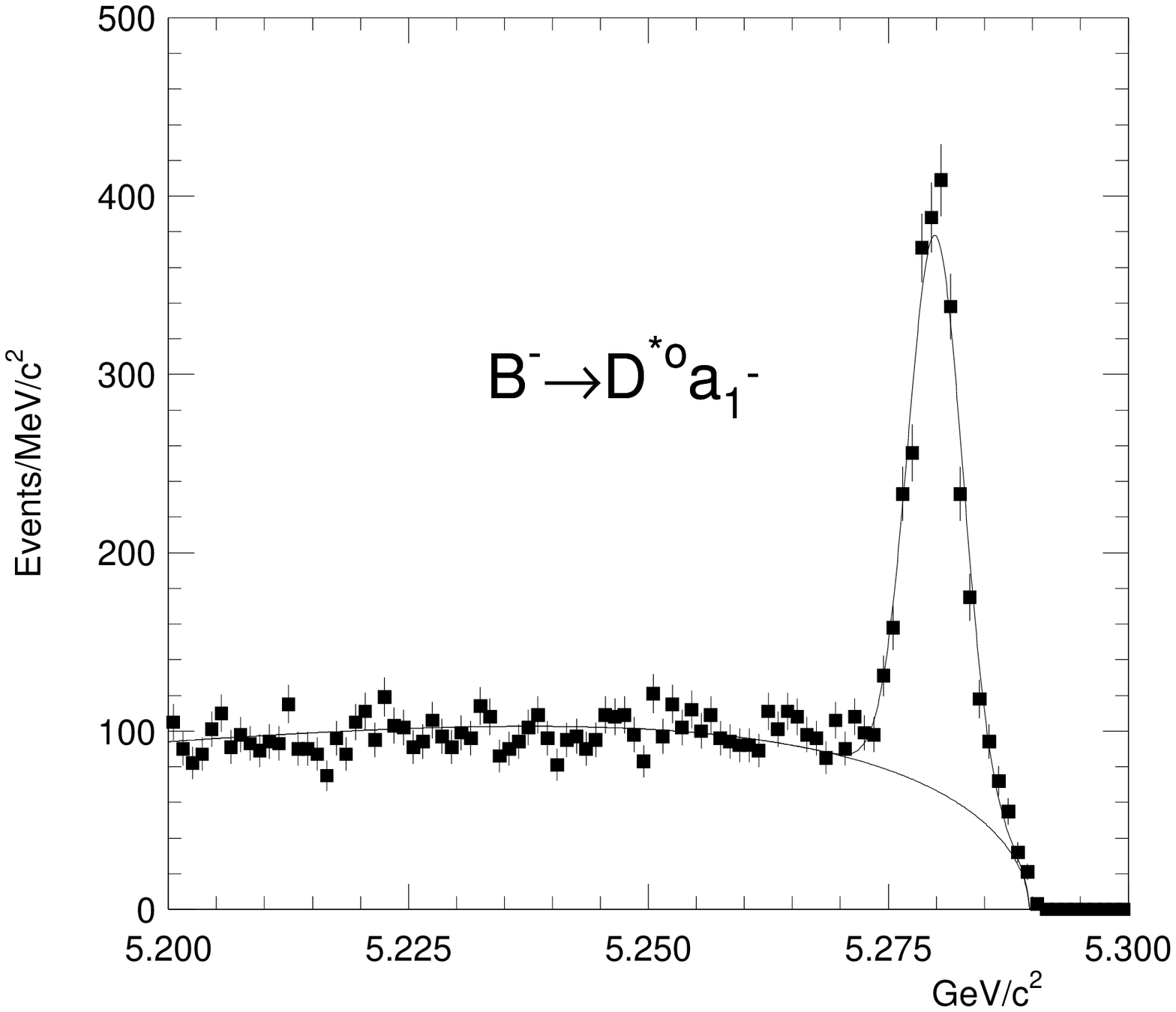}}      
\caption[]{
\label{LegsFigure6}
\small $m_{ES}$ distributions for $B^{-} \ra D^{(*)0}a_{1}^{-}$
with $|\Delta E|< 2.5 \sigma_{\Delta E}$.}
\end{figure}

\begin{figure}[ht]
\centerline{\epsfxsize 3.2 truein \epsfbox{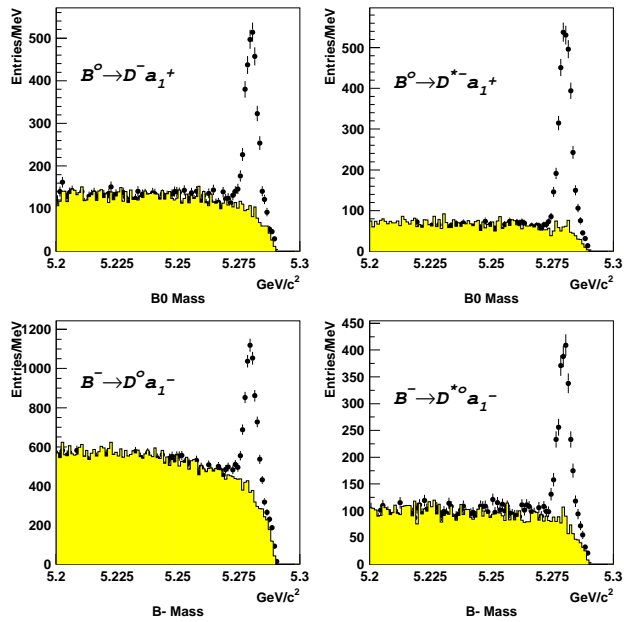}}
\caption[]{
\label{LegsFigure7}
\small  $m_{ES}$ distributions of $B \ra D^{(*)}a_{1}$ from data (dots) and
background modeled by Monte Carlo (shaded area).}
\end{figure}

The event selection efficiencies for each submode have been studied using
signal Monte Carlo data and are reported in Table 1. Estimates of the
statistical and systematic error of branching fraction measurements are
also listed in the table. The systematic error is mainly considered from 
the following sources (1) the branching fraction errors of intermediate 
decays, (2) the error of the integrated luminosity (about 1.1\%), (3) the 
error on the acceptance efficiency due to Monte Carlo statistics, and (4) the 
error on the charged track reconstruction efficiency, which is 1.2\% per 
charged track. An additional error of 1.6\% is added in quadrature to account 
for the uncertainty in the soft pion reconstruction efficiency.

\vbox{
\begin{table}
\caption{Event selection efficiencies and branching fraction error estimation.}
\begin{tabular}{lllcc} 
$B$ mode & $D$ mode& $\varepsilon_{MC}$ 
& $\Delta BF/BF$ (stat.) & $\Delta BF/BF$ (syst.) \\ \hline 
$\rule{0pt}{13pt} D^{*+}a_{1}^{-}$ & $K^{-}\pi^{+}$ & 9.9\% &  &  \\
 &           $K^{-}\pi^{+}\pi^{0}$ & 3.4\% & 3.4\% & 8.6\% \\
 &    $K^{-}\pi^{+}\pi^{+}\pi^{-}$ & 4.9\% &  &  \\  
\hline 
$\rule{0pt}{13pt} 
D^{+}a_{1}^{-}$& $K^{-}\pi^{+}\pi^{+}$ & 8.3\% & 3.0\% & 9.5\%\\
 &                   $K^{0}_{S}\pi^{+}$ & 8.0\% &  & \\ 
\hline 
$\rule{0pt}{13pt} D^{*0}a_{1}^{-}$& $K^{-}\pi^{+}$ & 5.9\% &  & \\
 &          $K^{-}\pi^{+}\pi^{0}$ & 2.0\% &  4.5\% & 11.5\%\\
 &   $K^{-}\pi^{+}\pi^{+}\pi^{-}$ & 2.7\% &  & \\ 
\hline
$\rule{0pt}{13pt} D^{0}a_{1}^{-}$& $K^{-}\pi^{+}$ & 8.8\% &  & \\
 &         $K^{-}\pi^{+}\pi^{0}$ & 4.8\% & 3.5\% & 9.8\%\\
 &  $K^{-}\pi^{+}\pi^{+}\pi^{-}$ & 5.3\% &  & \\ 
\end{tabular}
\end{table}

\begin{figure}[h]
\centerline{\epsfxsize 3.2 truein \epsfbox{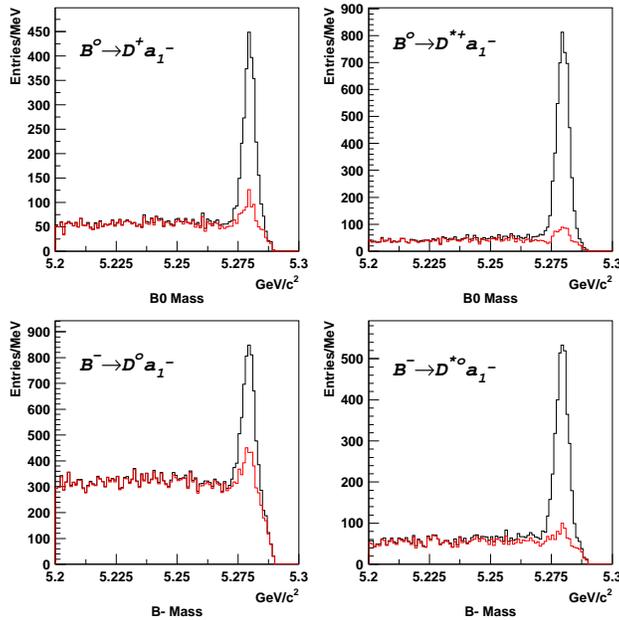}}
\caption[]{
\label{LegsFigure8}
\small Peaking background of non-resonant modes (dotted/colored line) underlays the 
signal of $B \ra D^{(*)}a_{1}$ ($B\bar{B}$ Monte Carlo data, with 
$m_{\rho^{0}\pi}< 4.0$ GeV/c$^{2}$).}
\end{figure}
}

\subsection{ Non-resonant modes $B \ra D^{(*)}\rho^{0}\pi$ and 
$B \ra D^{(*)}\pi\pi\pi$}

For the decays of $B \ra D^{(*)}\rho^{0}\pi$ and 
$B \ra D^{(*)}\pi\pi\pi$, the invariant masses of $\rho^{0}\pi$ and
$\pi\pi\pi$ are required to be less than 4.0 GeV/c$^{2}$ and center of 
mass momenta
of $\rho^{0}\pi$ and $\pi\pi\pi$ greater than 0.5 GeV/c. The $m_{ES}$ 
distributions for $B \ra D^{(*)}\rho^{0}\pi$ are shown in Fig. \,9. 

\begin{figure}[ht]
\centerline{\epsfxsize 3.2 truein \epsfbox{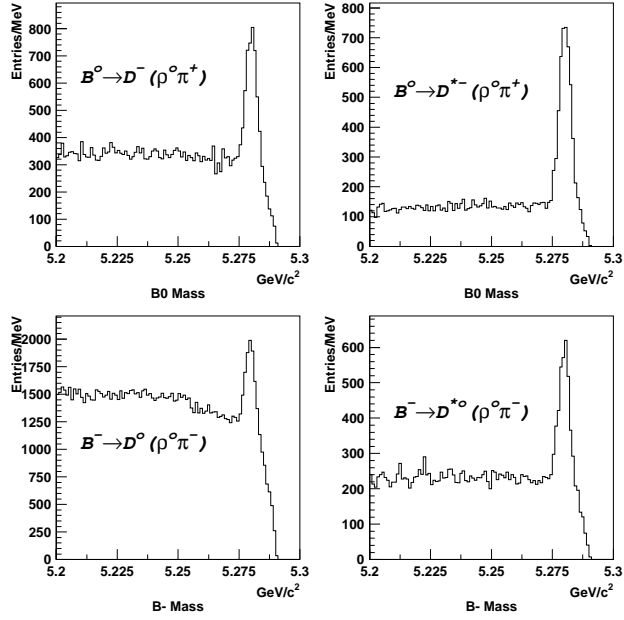}}
\caption[]{
\label{LegsFigure9}
\small Distribution of $m_{ES}$ for $B \ra D^{(*)}\rho^{0}\pi$. }
\end{figure}

The $B$ signal region of $B \ra D^{(*)}\rho^{0}\pi$, which is 
$|m_{ES}-5.280$ GeV/c$^{2}$ $|< 3\sigma_{m_{ES}}$ and 
$|\Delta E| <2.5 \sigma_{\Delta E}$ (see Fig. \,9) also contains  
$B \ra D^{(*)}a_{1}$ and $B \ra D^{(*)}\pi\pi\pi$ candidates. 
Monte Carlo study shows that the 
$\pi\pi\pi$ invariant mass 
spectra for $B \ra D^{(*)} \rho^{0} \pi^{-}$ and $B \ra D^{(*)} \pi^{-} 
\pi^{+} \pi^{-}$ are different. But when the cuts $|m_{\pi^{+}\pi^{-}}-0.77$
 GeV/c$^{2}$ $|< 0.15$ GeV/c$^{2}$ and center of mass momentum of 
$\pi^{-}\pi^{+}\pi^{-}$ greater than 0.5 GeV/c are applied to both modes, the
$\pi^{-}\pi^{+}\pi^{-}$ mass spectra for $B \ra D^{(*)} \rho^{0} \pi^{-}$ and 
$B \ra D^{(*)} \pi^{-} \pi^{+} \pi^{-}$ are almost the same, as
shown in Fig. \,10. So, we can use the $\pi^{-}\pi^{+}\pi^{-}$ spectra of 
$B \ra D^{(*)} \rho^{0} \pi$ to represent the spectra of 
$B \ra D^{(*)} \pi^{-} \pi^{+} \pi^{-}$. Monte Carlo also shows that the 
$\pi^{-}\pi^{+}\pi^{-}$ spectra of modes $B^{0} \ra D^{-} \rho^{0} \pi^{+}$, 
$B^{0} \ra D^{*-} \rho^{0} \pi^{+}$, $B^{-} \ra D^{0} \rho^{0} \pi^{-}$ and
$B^{-} \ra D^{*0} \rho^{0} \pi^{-}$ are quite similar because they have
similar topology and kinematics (see Fig. \,11).

\begin{figure}[ht]
\centerline{\epsfxsize 3.2 truein \epsfbox{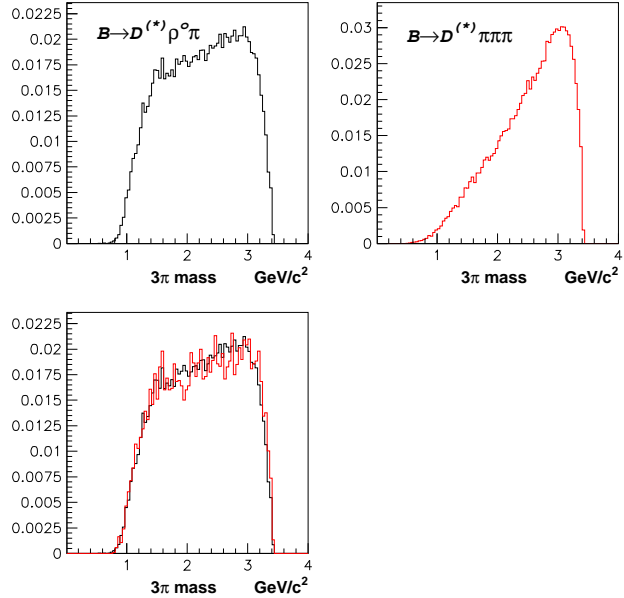}}
\vskip .1 cm
\caption[]{
\label{LegsFigure10}
\small Monte Carlo simulated distributions of $\pi^{-}\pi^{+}\pi^{-}$ mass 
for $B \ra D^{(*)}\rho^{0}\pi^{-}$ (upper left),
$B \ra D^{(*)} \pi^{-} \pi^{+} \pi^{-}$ (upper right) and 
$\pi^{-}\pi^{+}\pi^{-}$ mass (lower left) for $B \ra D^{(*)}\rho^{0}\pi^{-}$
(solid-line) and $B \ra D^{(*)} \pi^{-} \pi^{+} \pi^{-}$ (dotted/colored-line)
with the cuts $|m_{\pi^{+}\pi^{-}}-0.77$ GeV/c$^{2}$ $|<0.15 $ GeV/c$^{2}$ 
and center of mass momentum $p^{*}_{\pi^{-}\pi^{+}\pi^{-}} > 0.5$ GeV/c.
(Normalized distributions with arbitrary unit in the vertical axis).}
\end{figure}

\begin{figure}[ht]
\centerline{\epsfxsize 3.2 truein \epsfbox{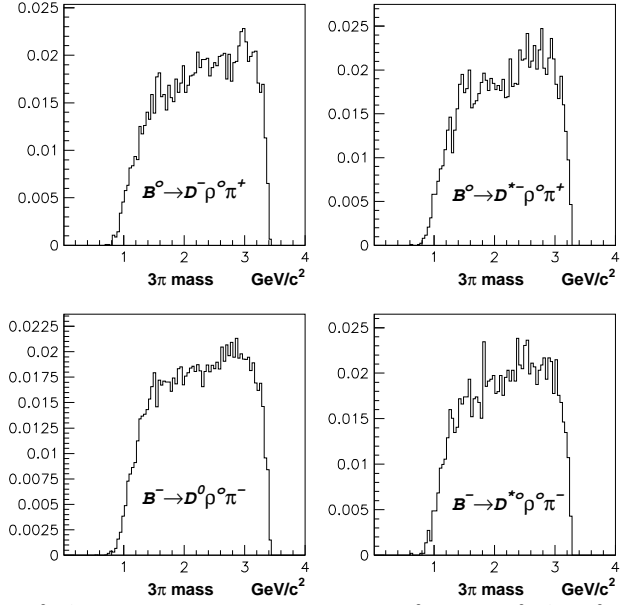}}
\vskip .1 cm
\caption[]{
\label{LegsFigure11}
\small Monte Carlo simulated
$\rho^{0}\pi^{+}$ mass distribution for decay 
$B^{0} \ra D^{-}\rho^{0}\pi^{+}$, 
$B^{0} \ra D^{*-}\rho^{0}\pi^{+}$,
$B^{-} \ra D^{0}\rho^{0}\pi^{-}$, and 
$B^{-} \ra D^{*0}\rho^{0}\pi^{-}$. 
(Normalized distributions with arbitrary unit in the vertical axis).}
\end{figure}

The invariant mass distributions of $\rho^{0} \pi^{+}$ in the $B$ signal 
region are shown in Fig. \,12. The shaded areas in the figures are the 
distributions from $m_{ES}$ sideband region
(5.22 GeV/c$^{2} < m_{ES} <$ 5.26 GeV/c$^{2}$), and were scaled with the 
ratio of the integral of the Argus function in the $m_{ES}$ sideband region 
to the integral in the $m_{ES}$ signal region.\\

To estimate the fraction of $a_{1}(1260)$ in the $\rho^{0} \pi^{-}$ mass 
distribution, the $\rho^{0} \pi^{-}$ mass distribution,
with the $m_{ES}$ sideband background subtracted, 
are fitted to a Breit-Wigner distribution as $a_{1}(1260)$ signal
plus the distribution of non-resonant $\rho^{0} \pi^{+}$ mass. The fits of 
$B^{0}$ modes are shown in Fig. \,13.
The non-resonant $\rho^{0} \pi^{+}$ mass distribution are from the 
reconstruction of the $\rho^{0} \pi^{+}$ mass in signal 
$B \ra D^{(*)} \rho^{0} \pi^{+}$ Monte Carlo data. 
It is found that the $a_{1}(1260)$ signal is dominant
in the mass spectra of $\rho^{0} \pi^{+}$ in both $B^{0}$ and $B^{-}$
decay; therefore, the decay of $B \ra D^{(*)}a_{1}(1260)$ 
is dominant in $B$ decay 
with final state $D^{(*)}\rho^{0}\pi$.
To separate the non-resonant mode $B \ra D^{(*)} \rho^{0} \pi$
and $B \ra D^{(*)}\pi\pi\pi$, a Dalitz plot analysis will be important.

\vbox{
\begin{figure}[ht]
\centerline{\epsfxsize 3.2 truein \epsfbox{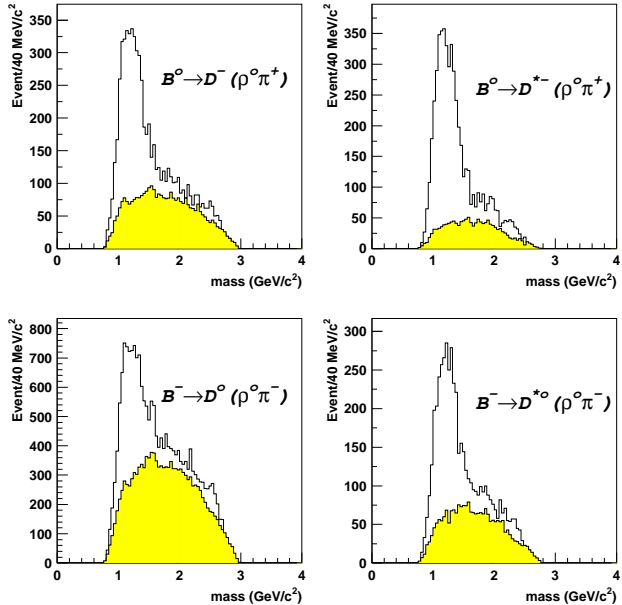}}
\caption[]{
\label{LegsFigure12}
\small $\rho^{0}\pi$ mass distribution in the $m_{ES}$ signal region with
the shaded area showing the $m_{ES}$ sideband background.}  
\end{figure}
}

\begin{figure}[ht]
\centerline{\epsfxsize 2.0 truein \epsfbox{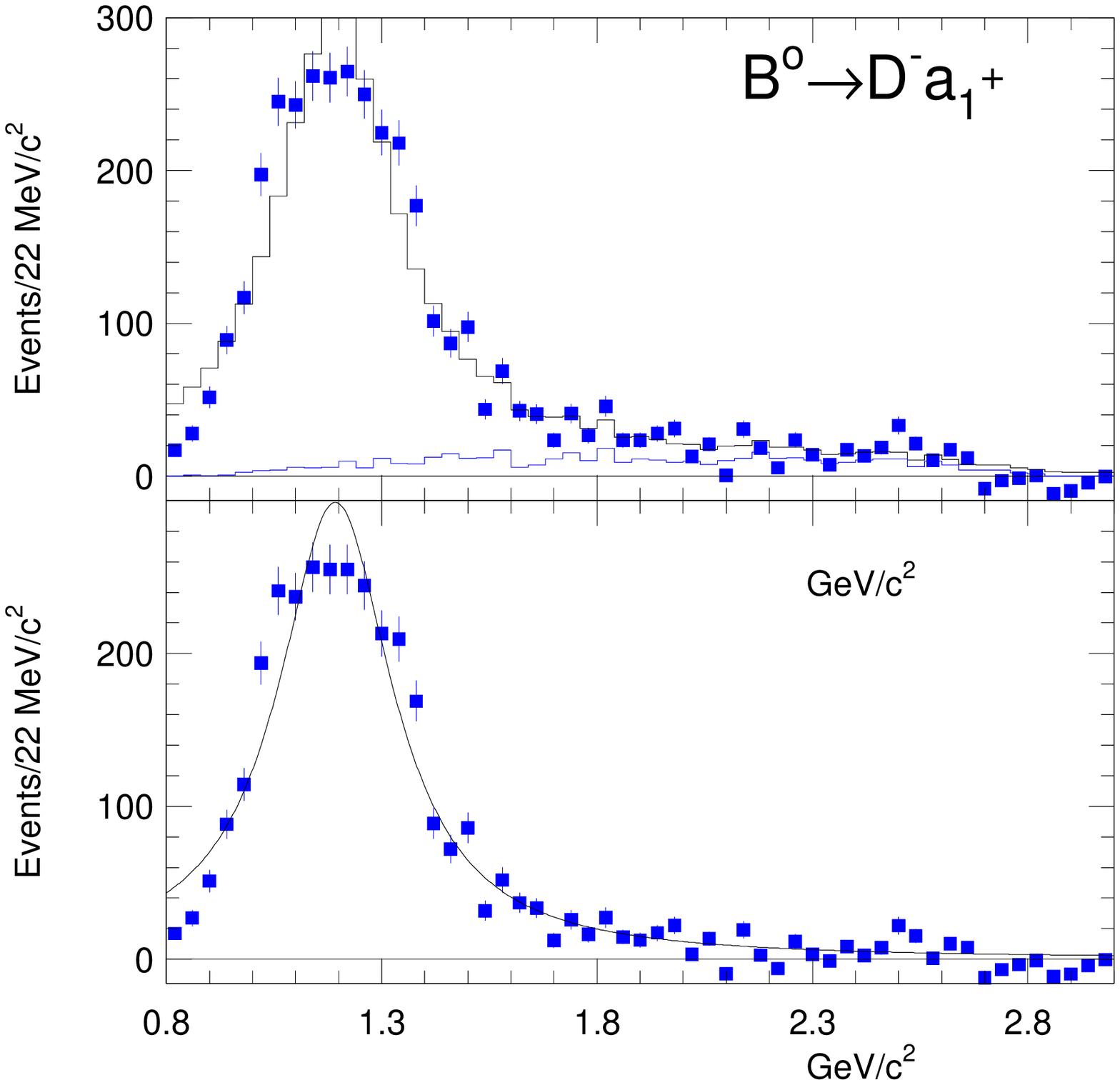}
\epsfxsize 2.0 truein \epsfbox{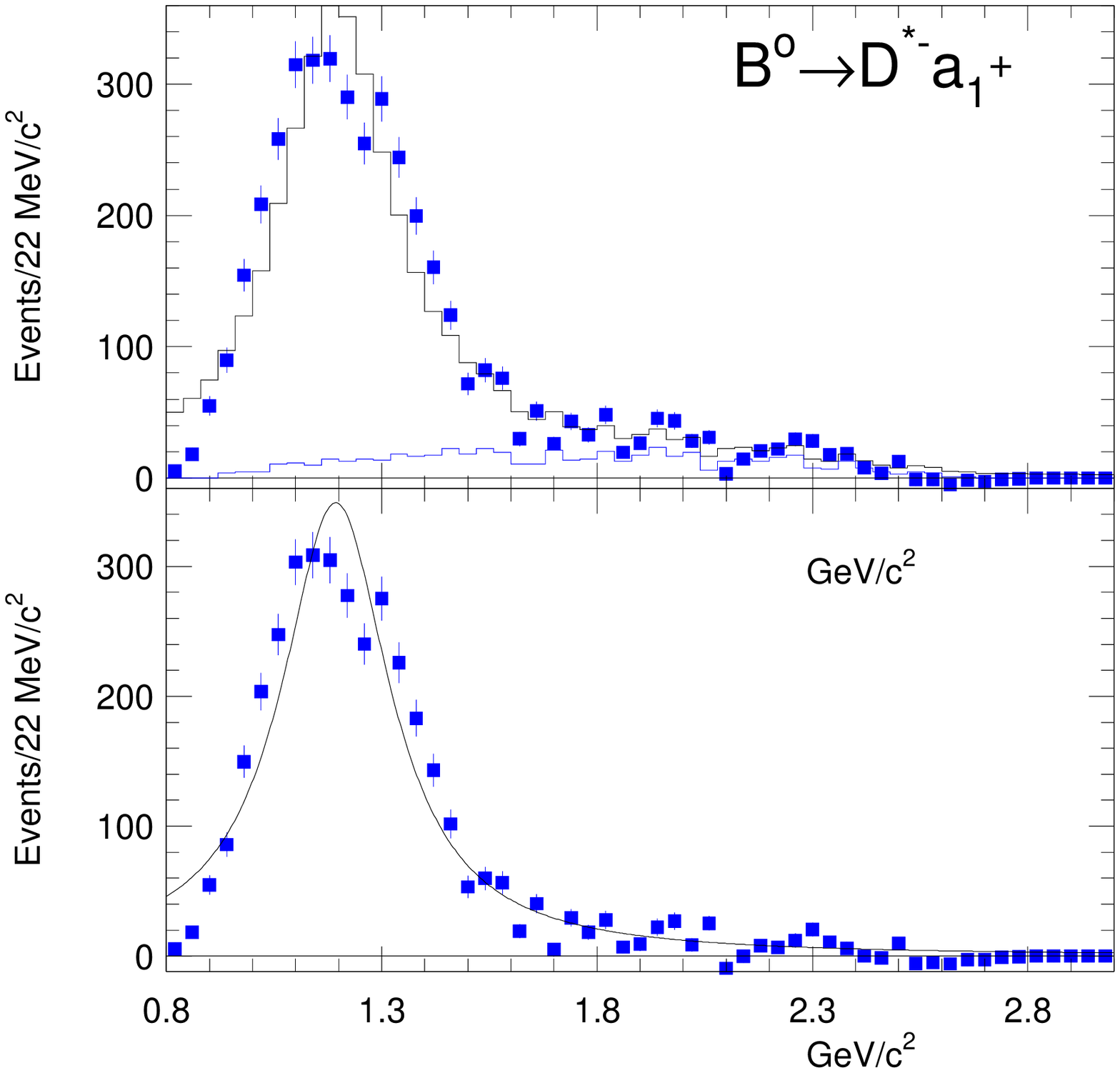}}
\caption[]{
\label{LegsFigure13}
\small Background-subtracted $\rho^{0}\pi$ mass spectra of data (dots) are 
       fitted to a Breit-Wigner function as $a_{1}(1260)$ signal, plus the 
       non-resonant $\rho^{0}\pi$ distribution as background. Lower plots 
       show the non-resonant component subtracted $a_{1}$ mass. }
\end{figure}

\vspace{-12pt}
\section{SUMMARY}
\vspace{-10pt}

The hadronic decay of $B \ra D^{(*)} a_{1}(1260)$ and non-resonant
modes $B \ra D^{(*)} \rho^{0} \pi$, $B \ra D^{(*)} \pi \pi \pi$ are
studied with a large data sample. The preliminary result shows 
that $B \ra D^{(*)} a_{1}(1260)$ is dominant in $B$ decays with the final 
state $D^{(*)}\rho^{0}\pi$. The errors on the branching fraction measurements 
are estimated and improved compared with previous measurements. Further study 
is ongoing and final results will be published soon.

\vspace{-3pt}
\section{ACKNOWLEDGMENTS}
\vspace{-10pt}

We are grateful for the excellent luminosity and machine conditions provided
by our PEP-II colleagues.
The collaborating institutions
wish to thank SLAC for its support and kind hospitality. This work is 
supported by DOE and NSF (USA), NSERC (Canada), IHEP (China), CEA and 
CNRS-IN2P3 (France), BMBF and DFG (Germany), INFN (Italy), FOM 
(The Netherlands), NFR (Norway), MIST (Russia), and PPARC (United Kingdom).
Individuals have received support from the A. P. Sloan Foundation, Research 
Corporation, and Alexander von Humboldt Foundation.

\end{document}